\pdfoutput=1
\documentclass[10pt]{article}
\usepackage{geometry}
\usepackage{xcolor}
\definecolor{graycolor}{gray}{0.9}
\usepackage{microtype}
\usepackage{setspace}
\usepackage[utf8]{inputenc}
\usepackage[english]{babel}
\usepackage{times}
\usepackage[numbers,sort&compress]{natbib}
\usepackage{natbib}

\usepackage{titlesec}
\titleformat {\section} [block] {\raggedright \fontsize{10}{10}\selectfont\bfseries} {\thesection. \space} {0pt} {}
\titlespacing {\section} {0pt} {12pt} {6pt}
\titleformat {\subsection} [block] {\raggedright \fontsize{10}{10}\selectfont\itshape} {\thesubsection .\space} {0pt} {}
\titlespacing {\subsection} {0pt} {12pt} {6pt}
\titleformat {\subsubsection} [block] {\raggedright \fontsize{10}{10}\selectfont} {\thesubsubsection .\space} {0pt} {}
\titlespacing {\subsubsection} {0pt} {12pt} {6pt}
\titleformat {\paragraph} [block] {\raggedright \fontsize{10}{10}\selectfont} {} {0pt} {}
\titlespacing {\paragraph} {0pt} {12pt} {6pt}

\usepackage{array} \newcommand{\PreserveBackslash}[1]{\let\temp=\\#1\let\\=\temp}
\newcolumntype{C}[1]{>{\PreserveBackslash\centering}m{#1}}
\newcolumntype{R}[1]{>{\PreserveBackslash\raggedleft}m{#1}}
\newcolumntype{L}[1]{>{\PreserveBackslash\raggedright}m{#1}}
\usepackage{lineno}
\usepackage{tabularx}
\usepackage{colortbl}
\usepackage{graphicx}
\usepackage{float}
\usepackage[export]{adjustbox}
\usepackage{caption}
\usepackage{fancyhdr}
\usepackage{amsmath}
\usepackage{lastpage}
\usepackage{layout}
\usepackage{setspace}
\usepackage{enumitem}
\usepackage{booktabs}
\usepackage{arydshln}
\usepackage{multirow}
\usepackage{color}
\usepackage{hyperref}
\hypersetup{
	colorlinks=true,
	linkcolor=blue,
	filecolor=blue,
	urlcolor=black,
	citecolor=cyan,
}

\setcitestyle{open={[},close={]},citesep={,\!},numbers}

\fancypagestyle{firstpage}{
    \setlength{\headsep}{2.2cm}
    
    \setlength{\footskip}{1.5cm}
    \fancyhf{}
    \lhead{\begin{table}[H]
        \centering
        \begin{tabular}{L{2.5cm}C{10cm}C{3.1cm}R{2cm}}
            \includegraphics[scale=0.035]{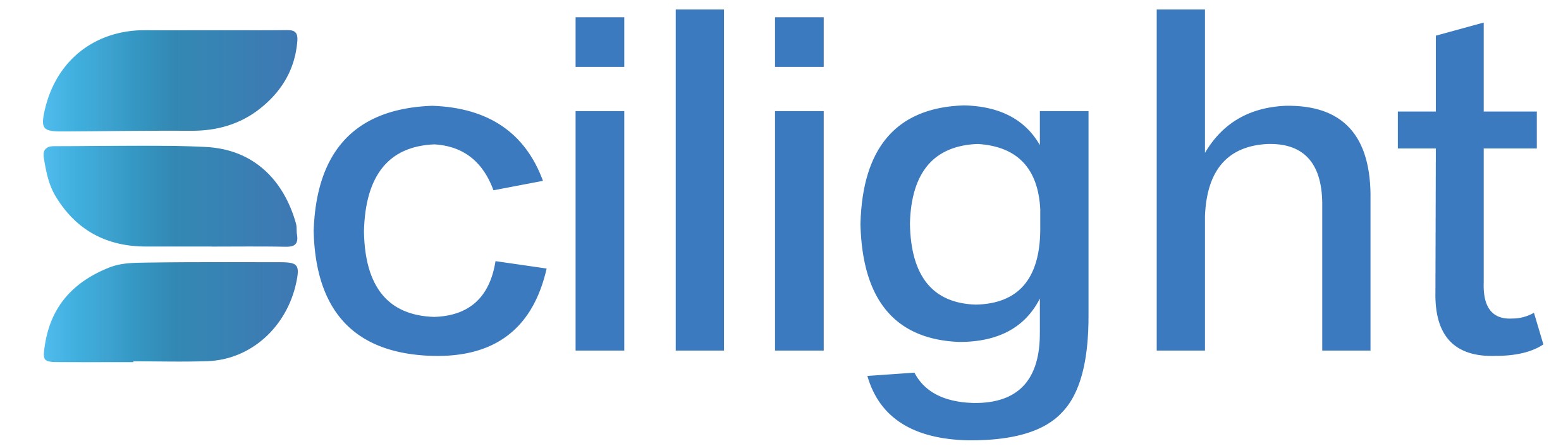} \vspace{-6pt}& \cellcolor{graycolor}\begin{tabular}[c]{@{}c@{}}\textit{International Journal of Gravitation and Theoretical Physics}\\ \href{https://www.sciltp.com/journals/ijgtp}{https://www.sciltp.com/journals/ijgtp}\end{tabular} & \includegraphics[scale=0.020]{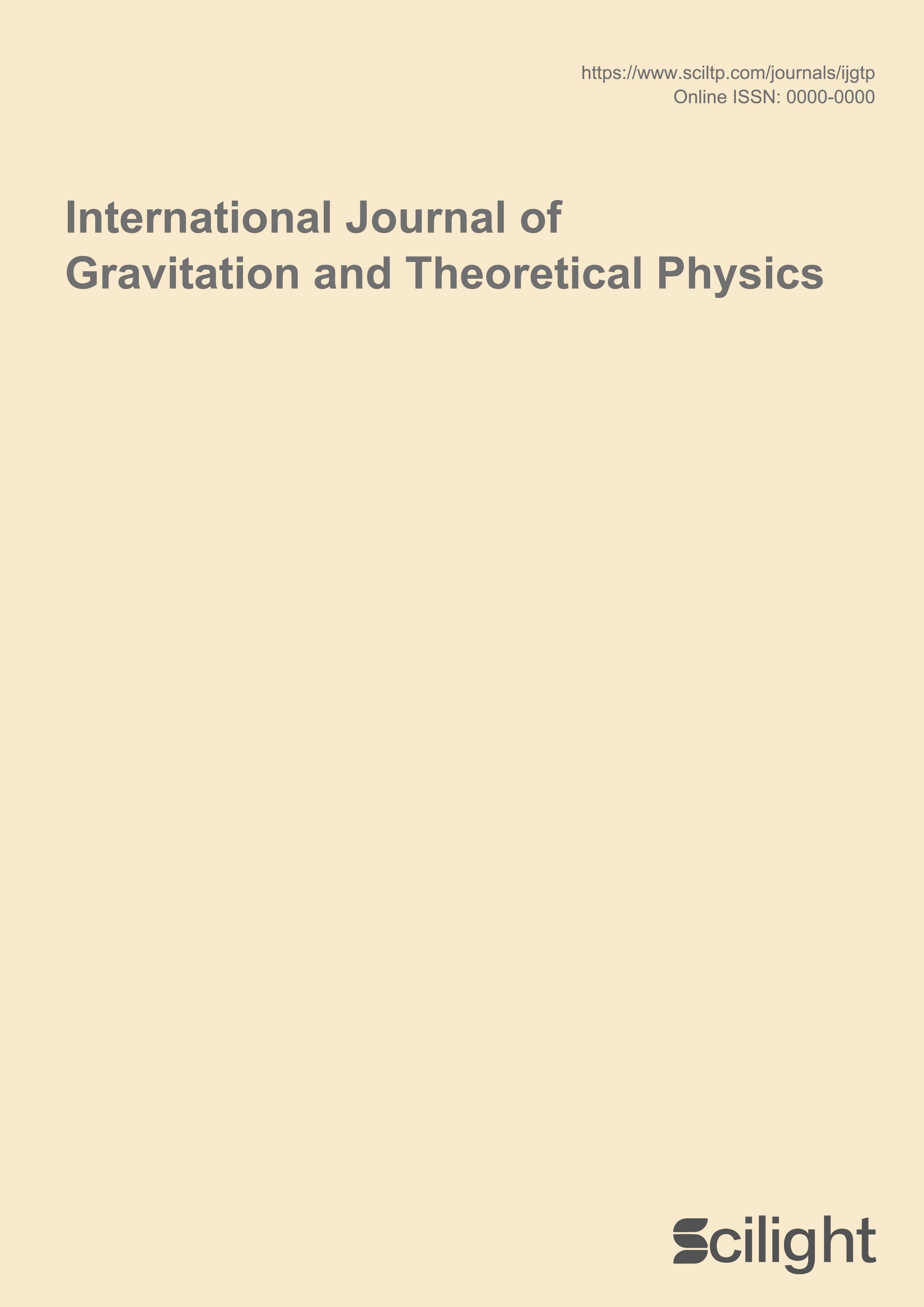} \vspace{-3pt}\\
        \end{tabular}
        \vspace{-22pt}
    \end{table}}
   
    \fancyfoot[C]{
        \vspace{-1.55cm}
        \begin{table}[H]
            \begin{minipage}[c]{0.15\columnwidth}
                \includegraphics[scale=0.5]{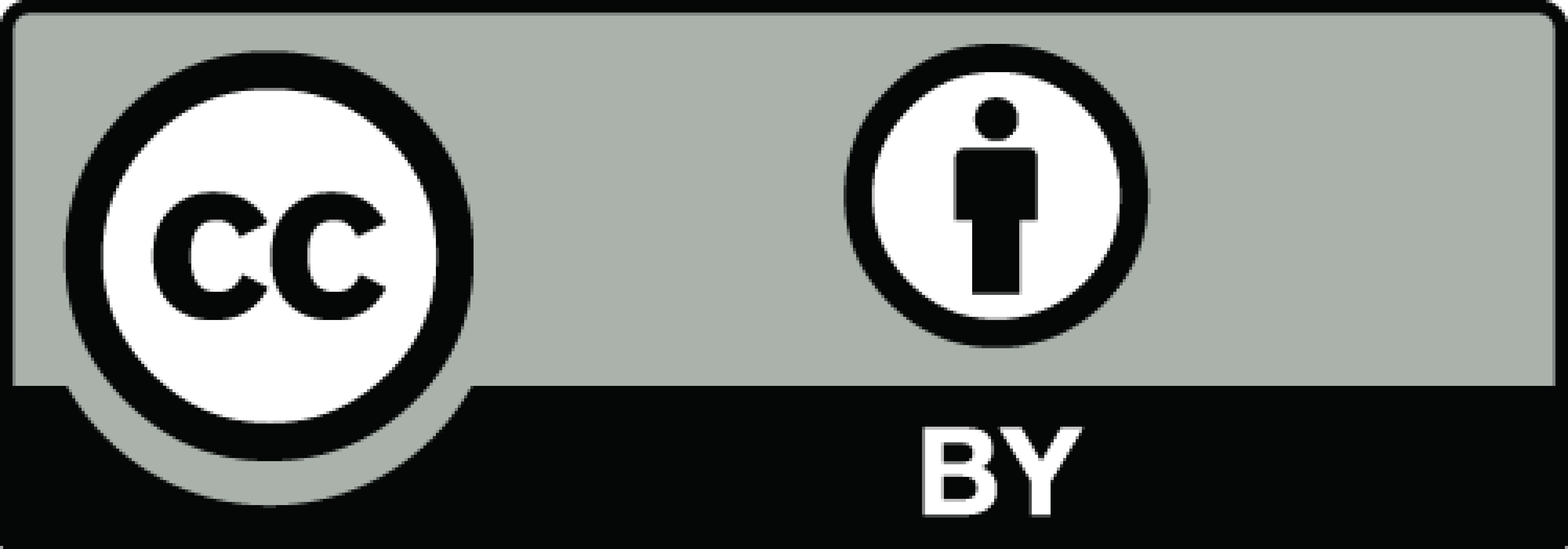} \vspace{1.1pt}
            \end{minipage}
            \hfill
            \begin{minipage}[c]{0.85\columnwidth}
                \scriptsize \textbf{Copyright:} © 2025 by the authors. This is an open access article under the terms and conditions of the Creative Commons Attribution (\mbox{CC BY}) license (\href{https://creativecommons.org/licenses/by/4.0/}{https://creativecommons.org/licenses/by/4.0/}). \\ \textbf{Publisher’s Note:} Scilight stays neutral with regard to jurisdictional claims in published maps and institutional affiliations.
            \end{minipage}
    \end{table}}
    \vspace{-0.55cm}
}

\def\imo{i}
\def\Order#1{{\cal O}\left(#1\right)}

\begin{document}
\newgeometry{left=2.5cm, right=2.5cm, top=1.8cm, bottom=4cm}
	\thispagestyle{firstpage}
	\nolinenumbers
	{\noindent \textit{Article}}
	\vspace{4pt} \\
	{\fontsize{18pt}{10pt}\textbf{Convergence of Higher-Curvature Expansions Near the \\Horizon: Hawking Radiation from Regular Black Holes}  }
	\vspace{16pt} \\
	{\large Roman. A. Konoplya \textsuperscript{1,*} and Alexander Zhidenko \textsuperscript{1,2} }
	\vspace{6pt}
	\begin{spacing}{0.9}
		{\noindent \small
			\textsuperscript{1}	Research Centre for Theoretical Physics and Astrophysics, Institute of Physics, Silesian University in Opava, \\ \hspace*{0.35em} Bezručovo náměstí 13, CZ-74601 Opava, Czech Republic \\
			\textsuperscript{2}	Centro de Matemática, Computação e Cognição (CMCC), Universidade Federal do ABC (UFABC), \\
            \hspace*{0.4em} Rua Abolição, Santo André 09210-180, SP, Brazil \\
		    {*}  \parbox[t]{0.98\linewidth}{Correspondence: roman.konoplya@gmail.com}\vspace{6pt}\\
		\footnotesize	\textbf{How To Cite}: Konoplya, R.A.; Zhidenko, A. Convergence of Higher-Curvature Expansions Near the Horizon: Hawking Radiation from Regular Black Holes. \emph{International Journal of Gravitation and Theoretical Physics} \textbf{2025}, \emph{1}(1), 5. 
        }\\
	\end{spacing}

\begin{table}[H]
\noindent\rule[0.15\baselineskip]{\textwidth}{0.5pt}
\begin{tabular}{lp{12.4cm}}
 \small
  \begin{tabular}[t]{@{}l@{}}
  \footnotesize  Received: 14 May 2025  \\
  \footnotesize  Revised: 10 July 2025  \\
  \footnotesize  Accepted: 23 July 2025 \\
  \footnotesize  Published: 25 July 2025
  \end{tabular} &
  \textbf{Abstract:} A recently proposed model incorporating a series of higher-curvature corrections allows for analytic black-hole solutions at each order of the expansion, with a fully regular black hole emerging in the limit of infinite number of terms. An important question that arises within this framework is how rapidly the series converges. For those classical observables, which are primarily determined by the geometry near the peak of the effective potential, it has been previously shown that the series converges remarkably fast, often within the first two orders. However, this rapid convergence does not extend to quantities such as Hawking radiation, which are highly sensitive to the geometry near the event horizon. Although each successive order yields a result that is significantly closer to that of the full infinite series, several terms are typically required to obtain a sufficiently accurate approximation of the regular black hole in this context.
  \\
  \\
  &
  &

\vspace{4pt}


\mbox{}\vspace{-12pt}\\

\end{tabular}
\noindent\rule[0.15\baselineskip]{\textwidth}{0.5pt}
\end{table}

\section{Introduction}

The problem of resolving black-hole singularities has motivated the exploration of gravitational theories beyond general relativity (GR). Among these, higher-curvature corrections to the Einstein–Hilbert action provide a natural path for ultraviolet completions of GR. In this context, a particularly promising framework is offered by the class of higher-dimensional quasi-topological gravities \cite{Oliva:2010eb,Myers:2010ru,Dehghani:2011vu,Ahmed:2017jod,Cisterna:2017umf}, which enable the construction of static, spherically symmetric black hole solutions through analytic methods at every order in the curvature. Recently, it was demonstrated that an infinite series of such corrections leads to a fully regular black hole geometry without requiring any additional matter fields~\cite{Bueno:2024dgm}. Unlike earlier regular black hole models based on exotic matter content, this approach achieves singularity resolution purely within the gravitational sector, with a controlled effective field theory interpretation.

A recent breakthrough in higher-curvature gravity shows that an infinite tower of quasi-topological corrections can dynamically lead to the formation of regular black holes during gravitational collapse~\cite{Bueno:2024eig}. Unlike earlier ad hoc or matter-supported constructions, this approach provides a singularity-free, geodesically complete spacetime purely from modified gravitational dynamics, and admits an analytic treatment of thin-shell collapse within a fully consistent theory \cite{Bueno:2024zsx}.

A key advantage of this construction is that it admits analytic black-hole solutions at each order of the curvature expansion. This allows a systematic study of the convergence behavior of physical observables computed in the truncated theories toward those in the full (resummed) geometry. In particular, classical observables that probe the region near the peak of the effective potential---such as quasinormal frequencies or photon sphere characteristics---were found in~\cite{Konoplya:2024hfg} to converge rapidly, often already at second order in the curvature. However, quantum processes like Hawking radiation, which are sensitive to the near-horizon geometry, exhibit a slower convergence pattern. This raises the important question of how many terms in the curvature expansion are required to accurately approximate the semiclassical radiation spectrum of the regular black hole.

In this work, we address this issue by comparing the grey-body factors and Hawking radiation emission rates for black holes arising in the infinite-series higher-curvature theory with those in the truncated theories. Our analysis quantifies how the radiation spectra computed from finite-order truncations approach the full-theory result, and identifies the minimum number of terms required to reproduce the features of the regular solution with reasonable~accuracy.

The paper is organized as follows. In Section~\ref{sec:equations}, we briefly review the black-hole solutions arising in a gravitational theory with an infinite series of higher-curvature corrections, focusing on the structure of the metric and its truncation at finite order. In Section~\ref{sec:Hawking}, we describe the numerical method used to compute grey-body factors and Hawking radiation, including the near-horizon expansion and asymptotic fitting procedures. Section~\ref{sec:results} presents a comparative analysis between the full regular black hole and its finite-order approximations, highlighting the convergence behavior of physical observables such as the Hawking temperature and radiation spectrum. Finally, Section~\ref{sec:conclusions} summarizes our findings and discusses their implications for the viability of truncated higher-curvature models in capturing quantum near-horizon effects.
	
\section{Black Holes in a Theory with Infinite Tower of Higher Curvature Corrections}\label{sec:equations}
	
\textls[-15]{Following \cite{Bueno:2024dgm}, the action of the Einsteinian theory with $N$ terms of the higher-curvature corrections has the form}
\begin{equation}\label{QTaction}
I_{\rm QT}=\frac{1}{16\pi G} \int \mathrm{d}^Dx \sqrt{|g|} \left[R+\sum_{n=2}^{N+1} \alpha_n \mathcal{Z}_n \right]\, ,
\end{equation}
where $\alpha_n$ are arbitrary coupling constants with dimensions of length$^{2(n-1)}$ and $\mathcal{Z}_n$ are the quasi-topological densities \cite{Bueno:2019ycr}. These higher-curvature terms allow for analytic black hole solutions at each order of the expansion.

The spherically symmetric $D$-dimensional black hole is given by the following metric:
\begin{eqnarray}\label{metric}
ds^2 &=& - f(r) dt^2 + \frac{dr^2}{f(r)} + r^2 d\Omega^2_{D-2} \, ,
\\\nonumber &&f(r)\equiv1-r^2\psi(r)\, ,
\end{eqnarray}
such that $\psi(r)$ satisfies
\begin{equation}\label{hequation}
h(\psi(r)) = \frac{\mu}{r^{D-1}}\, ,
\end{equation}
where $\mu$ is the positive constant proportional to the ADM mass. This functional form of $h(\psi)$ arises naturally from the field equations associated with the action \eqref{QTaction}, as the following sum:\vspace{-6pt}
\begin{equation}\label{hseries}
h(\psi) \equiv \psi + \sum_{n=2}^{N+1} \alpha_n \psi^n\, .
\end{equation}
\vspace{-10pt}

When all the coupling coefficients vanish, the metric reduces to the $D$-dimensional generalization of the vacuum Einstein solution, known as the Tangherlini metric~\cite{Tangherlini:1963bw}. Hawking radiation from those higher-dimensional black holes without taking into account higher-curvature corrections, has been extensively studied in the context of various brane-world models~\cite{Kanti:2004nr}, where the black hole's size is assumed to be much smaller than the size of the extra dimensions~\cite{Antoniadis:1990ew,Arkani-Hamed:1998jmv,Randall:1999ee}.

The values $\alpha_n$ are arbitrary constants. Two simple ad hoc choices are \cite{Bueno:2024dgm}
\begin{subequations}\label{Haywardchoice}
\begin{align}
\alpha_n&=\alpha^{n-1},\\
\alpha_n&=\dfrac{\alpha^{n-1}}{n},
\end{align}
\end{subequations}
leading to the definition of $h(\psi)$ through partial sum of a series,
\vspace{-12pt}

\begin{subequations}\label{hseriesHayward}
\begin{align}
h(\psi) &\equiv \displaystyle \sum_{n=1}^{N+1} \alpha^{n-1} \psi^n+\Order{\alpha}^{N+1}\,,\\
h(\psi) &\equiv \displaystyle \sum_{n=1}^{N+1} \dfrac{\alpha^{n-1}}{n} \psi^n+\Order{\alpha}^{N+1}\,.
\end{align}
\end{subequations}

This definition means that we neglect the terms of the order $\alpha^{N+1}$ and higher, which correspond to the higher-curvature corrections to the theory (\ref{QTaction}). Such a truncation allows us to examine how well a finite-order theory approximates geometry of the full (non-truncated) theory.

The series (\ref{hseriesHayward}) converges to
\begin{subequations}\label{Haywardh}
\begin{align}
\displaystyle \lim_{N\to\infty}h(\psi)&=\dfrac{\psi}{1-\alpha \psi}\,,\\
\displaystyle \lim_{N\to\infty}h(\psi)&=-\dfrac{\ln(1-\alpha \psi)}{\alpha}\,.
\end{align}
\end{subequations}

It follows from Equation (\ref{hequation}) that the metric function can be represented as a series expansion with respect to $\alpha$,
\begin{subequations}\label{fseriesHayward}
\begin{align}
f(r) &=  \displaystyle  1 + r^2 \sum_{n=1}^{N+1}\alpha^{n-1} \left(\dfrac{-\mu}{r^{D-1}}\right)^n+\Order{\alpha}^{N+1}\,,\\
f(r) &=  \displaystyle  1 + r^2 \sum_{n=1}^{N+1}\dfrac{\alpha^{n-1}}{n} \left(\dfrac{-\mu}{r^{D-1}}\right)^n+\Order{\alpha}^{N+1}\,,
\end{align}
\end{subequations}
which approaches to the metric function with no singular points, yielding a fully regular black-hole geometry,
\begin{subequations}\label{Haywardmetric}
\begin{align}
\displaystyle \lim_{N\to\infty}f(r) &= 1-\dfrac{\mu r^2}{r^{D-1}+\alpha\mu}, \\
\displaystyle \lim_{N\to\infty}f(r) &= 1-\dfrac{r^2}{\alpha}\left(1-\exp\left(-\dfrac{\alpha\mu}{r^{D-1}}\right)\right).
\end{align}
\end{subequations}

The metric (a) is the $D$-dimensional generalization of the Hayward black hole.

In order to test the approximation by a finite sum, which corresponds to the approximated theory, which takes into account only a finite number of terms $N$, we compare the metric (\ref{fseriesHayward}) and the metric (\ref{Haywardmetric}), which corresponds to the infinite tower of higher-order curvature corrections.

For illustrative purposes, we restrict our analysis to the emission of scalar fields, assuming that the convergence properties for fields with higher spin do not differ significantly.
The general relativistic equations for a scalar field,
\begin{equation}\label{KGg}
\frac{1}{\sqrt{-g}}\partial_\mu \left(\sqrt{-g}g^{\mu \nu}\partial_\nu\Phi\right)=0,
\end{equation}
after separation of variables can be reduced to the wavelike form
\begin{equation}\label{wavelike}
\frac{d^2\Psi}{dr_*^2}+(\omega^2-V(r_*))\Psi(r_*)=0,
\end{equation}
where the tortoise coordinate is defined as follows:
\begin{equation}\label{tortoise}
dr_*=\frac{dr}{f(r)}.
\end{equation}

The effective potential is
\begin{equation}\label{potential}
V(r) = f(r)\left(\frac{\ell(\ell+D-3)}{r^2} + \frac{(D-2)(D-4)}{4r^2}f(r) + \frac{D-2}{2r}\frac{df}{dr}\right).
\end{equation}

\begin{figure}[H]
\centerline{\resizebox{0.95\linewidth}{!}{\includegraphics{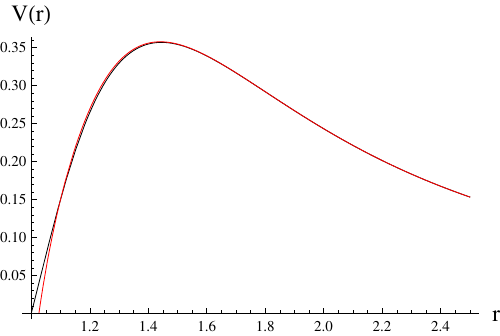}~~~\includegraphics{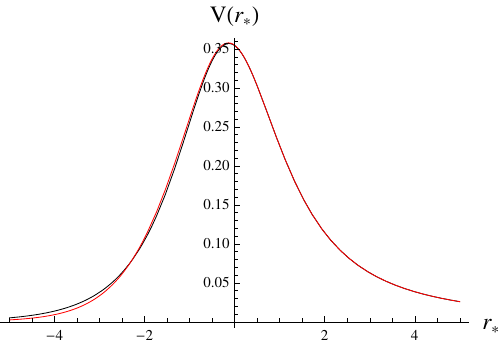}}}
\caption{Effective potential of the scalar field ($\ell=0$) for the $D=5$ Hayward black hole with $\mu=4/3$ and $\alpha=1/4$, which corresponds to $r_0=1$ (upper, black), vs. the effective potential for the truncated metric at $N=2$, which corresponds to $r_0\approx1.02616$ (lower, red). (\textbf{Left panel}): the potentials in terms of the radial coordinate. (\textbf{Right panel}): the potentials as functions of the corresponding tortoise coordinate.}\label{fig:D5Haywardpotentials}
\end{figure}

The effective potential has a form of potential barrier with a single peak (see Figure \ref{fig:D5Haywardpotentials}) and defines a standard scattering problem.

\vspace{-26pt}

\section{Grey-Body Factors and Hawking Radiation}\label{sec:Hawking}

Hawking radiation in higher-curvature corrected theories exhibits remarkable differences compared to higher dimensional general relativity, as even geometrically small corrections can lead to a significant---often orders of magnitude---suppression of the radiation and consequently longer-lived black holes~\cite{Konoplya:2010vz}.

In order to calculate grey-body factors via the numerical integration, we introduce a new function that is regular at the event horizon $r_0$:
\begin{equation}
P(r) = \Psi(r) \left(1 - \frac{r_0}{r} \right)^{\displaystyle \imo\omega/4\pi T_H},
\end{equation}
where $T_H$ is the Hawking temperature defined in a standard way,
\begin{equation}\label{Hawkingtemperature}
T_H=\frac{f'(r_0)}{4\pi}.
\end{equation}

By fixing the integration constant such that $P(r_0) = 1$, we expand Equation (\ref{wavelike}) near the horizon to determine $P'(r_0)$, thereby fully specifying the initial conditions for numerical integration. We then integrate Equation (\ref{wavelike}) numerically from the event horizon $r_0$ to a distant point $r_f \gg r_0$, and fit the resulting solution in the asymptotic region using the ansatz:
\begin{equation}\label{fit}
P(r) = Z_{\text{in}} P_{\text{in}}(r) + Z_{\text{out}} P_{\text{out}}(r),
\end{equation}
where the basis functions $P_{\text{in}}(r)$ and $P_{\text{out}}(r)$ represent incoming and outgoing waves, respectively, with the leading-order behavior at large $r$ given by:
\begin{eqnarray}
P_{\text{in}}(r) &\propto& e^{-\imo\omega r}, \nonumber\\
P_{\text{out}}(r) &\propto& e^{\imo\omega r}. \nonumber
\end{eqnarray}

This fitting procedure yields the coefficients $Z_{\text{in}}$ and $Z_{\text{out}}$. To ensure the accuracy of the result, we systematically increase the internal precision of the numerical integration, extend the value of $r_f$, and include more terms in the asymptotic expansions of $P_{\text{in}}(r)$ and $P_{\text{out}}(r)$. The stability of $Z_{\text{in}}$ and $Z_{\text{out}}$ under these changes confirms the reliability of the computation.

Once the coefficients $Z_{\text{in}}$ and $Z_{\text{out}}$ are determined, the absorption probability (grey-body factor) can be computed as follows:
\begin{equation}\label{absorbtion}
|{\cal A}_{\ell}|^2=1-|Z_{out}/Z_{in}|^2.
\end{equation}

Finally, using numerically found values of ${\cal A}_{\ell}$ we calculate the energy emission rates \cite{Harris:2003eg},
\begin{equation}\label{energy-emission}
\frac{dE}{dt} = \sum_{\ell=0}^\infty N_{\ell} \intop_0^{~\infty}{ \left| {\cal A}_{\ell} \right|^2 \frac{\omega}{\exp (\omega /T_H ) - 1}\frac{d\omega }{2\pi }},
\end{equation}
where the multiplicity factor for a scalar field in a higher-dimensional spacetime is
\begin{equation}\label{multiplicity}
N_{\ell}=\frac{(2 \ell + D - 3) (\ell + D - 4)!}{\ell! (D - 3)!}.
\end{equation}

It is important to note that the emission of particles with higher multipole numbers is exponentially suppressed. This suppression arises due to the presence of a larger centrifugal barrier in the effective potential for higher values of $\ell$, which reduces the transmission probability through the black hole's potential barrier. As a result, in practical calculations it is sufficient to include only the contributions from particles with the lowest few values of $\ell$, since higher-multipole modes contribute negligibly to the total flux. In the limit of vanishing higher-curvature corrections and for the $D = 4$ case, we recover the grey-body factors and emission rates of the Schwarzschild black hole, both for scalar fields and for fields of the Standard Model~\cite{Page:1976df,Page:1976ki}.

\section{Finite vs. Infinite Tower of Higher-Order Curvature Corrections}\label{sec:results}

\begin{figure}[H]
\centerline{\resizebox{0.75\linewidth}{!}{\includegraphics{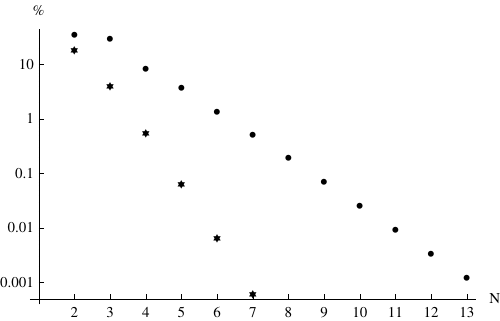}}}
\caption{Relative difference between the Hawking temperature obtained at the $N$-th order expansion and that for the full metric: Circles (top) for the metric (a) $D=5$, $\alpha=3\mu/16$ ($\mu T_H^2=1/12\pi^2\approx0.00844343$) and stars (bottom) for the metric (b) $D=6$, $\alpha^{3/2}=\mu/4\sqrt{2}\ln(2)$ ($\mu T_H^3=(5\ln2-2)^2\ln2/32\pi^3\approx0.00219985$).}\label{fig:THdev}
\end{figure}

Here we compare the energy-emission rates from the black hole of the same asymptotic mass $\mu$ given by finite number and by the infinite sum of higher-order curvature corrections, that is, by the truncated metric functions (\ref{fseriesHayward}) and the full regular metric function (\ref{Haywardmetric}). It is clear that, once we fix values of $\mu$ and $\alpha$, the value of the event horizon radius $r_0$, which can be obtained by solving the equation
\begin{equation}\label{r0eq}
f(r_0)=0,
\end{equation}
differs for different orders $N$ in (\ref{fseriesHayward}) and (\ref{Haywardmetric}). As a consequence, the theory with finite number of the curvature corrections, gives only approximate values for the radius of the event horizon of the full metric. However, the radius of the event horizon is not a gauge invariant observable, because it depends on the coordinate system  under consideration. Therefore we will study Hawking temperature (\ref{Hawkingtemperature}), which is a gauge invariant quantity defined at the horizon. In Figure \ref{fig:THdev} we illustrate the convergence of the Hawking temperature with respect to the number of the correction terms $N$.

\begin{figure}[H]
\centerline{\resizebox{0.75\linewidth}{!}{\includegraphics{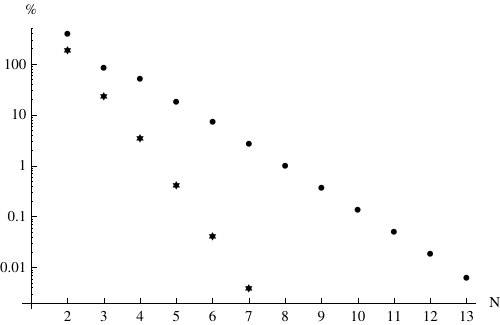}}}
\caption{Relative difference between the scalar particles energy-emission rate obtained at the $N$-th order expansion and that for the full metric: Circles (top) for the metric (a) $D=5$, $\alpha=3\mu/16$ and stars (bottom) for the metric (b) $D=6$, $\alpha^{3/2}=\mu/4\sqrt{2}\ln(2)$. In the limit of regular black holes we have, correspondingly, $\mu\frac{dE}{dt}\approx0.00005503$ (Hayward) and $\mu^{2/3}\frac{dE}{dt}\approx0.00005606$.}\label{fig:EEratedev}
\end{figure}

In Figure \ref{fig:EEratedev}, we present the relative difference between the energy emission rates of a scalar field computed using the truncated theories and that for the full theory. Hawking temperature is the main parameter governing the emission rate. While the dominant source of this difference arises from the deviation of the Hawking temperature from its value for the limit of regular black hole, there is an additional deviation due to the difference in grey-body factors for the truncated and full theory. The latter arises from the fact that the effective potential used in the truncated theory differs from that of the full theory (see Figure \ref{fig:D5Haywardpotentials}) in the near-horizon zone, leading to discrepancies in the transmission probabilities.

\vspace{-16pt}

\begin{figure}[H]
\centerline{\resizebox{0.95\linewidth}{!}{\includegraphics{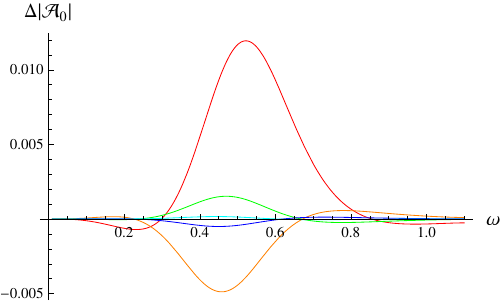}\includegraphics{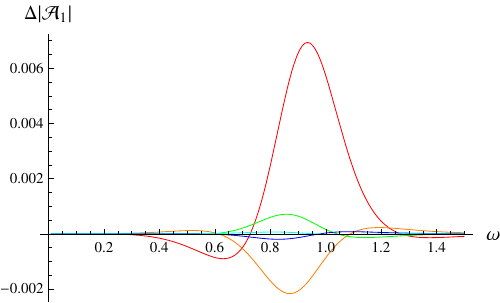}}}
\caption{\textls[-15]{Difference between the grey-body factors of scalar field for the $D=5$ Hayward black hole with $\mu=4/3$ and $\alpha=1/4$ and the expansion of the order 2 (red), 3 (orange), 4 (green), 5 (blue), and 6 (cyan) for $\ell=0$ (\textbf{left panel}) and $\ell=1$ (\textbf{right panel}).}}\label{fig:GBdevd5}
\end{figure}

\textls[-15]{It is important to note that the grey-body factors, which characterize the transmission through the potential barrier, converge more rapidly than the Hawking temperature in the truncated theory. As shown in Figure \ref{fig:GBdevd5}, for the $D = 5$ Hayward black hole, the sixth-order expansion yields transmission probabilities that differ negligibly from those in the full theory. Moreover, the convergence improves further for higher multipole numbers $\ell$. Therefore, the principal reason of slow convergence of the emission rates comes from the slow convergence  of the Hawking~temperature.}

\vspace{-16pt}

\begin{figure}[H]
\centerline{\resizebox{0.95\linewidth}{!}{\includegraphics{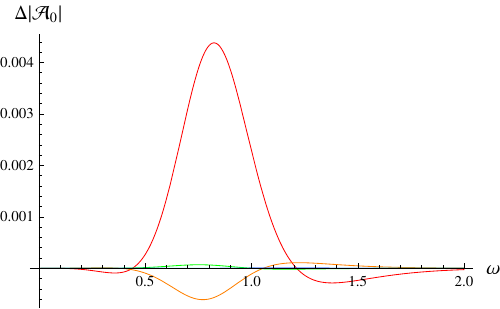}\includegraphics{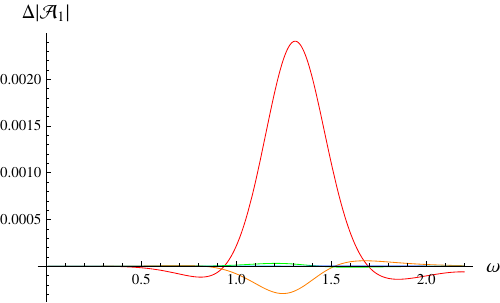}}}
\caption{Difference between the grey-body factors of scalar field for the $D=6$ black hole (b) with $\mu=2\ln2$ and $\alpha=1/2$ and the expansion of the order 2 (red), 3 (orange), 4 (green), 5 (blue), and 6 (cyan) for $\ell=0$ (\textbf{left panel}) and $\ell=1$ (\textbf{right panel}).}\label{fig:GBdevd6}
\end{figure}

\begin{figure}[H]
\centerline{\resizebox{0.9\linewidth}{!}{\includegraphics{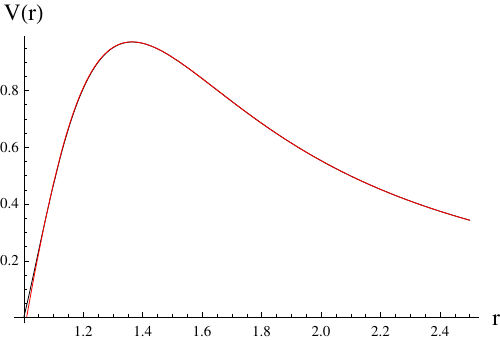}~~~\includegraphics{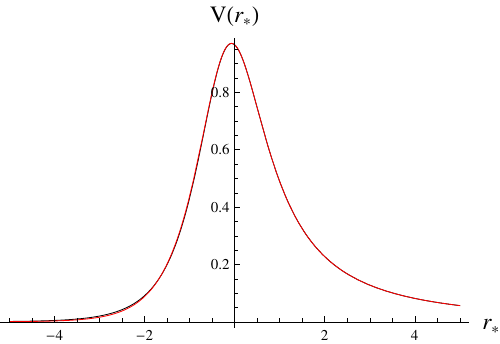}}}
\caption{\textls[-15]{Effective potential of the scalar field ($\ell=0$) for the $D=6$ black hole (b) with $\mu=2\ln2$ and $\alpha=1/2$, which corresponds to $r_0=1$ (upper, black), vs. the effective potential for the truncated metric at $N=2$, which corresponds to $r_0\approx1.00964$ (lower, red). (\textbf{Left panel}): the potentials in terms of the radial coordinate. (\textbf{Right panel}):~the potentials as functions of the corresponding tortoise coordinate.}}\label{fig:D6potentials}
\end{figure}

The convergence of the grey-body factors for a black hole (b) is faster than that for (a) (at least for the chosen values of the parameters) (see Figure \ref{fig:GBdevd6}),  because the potentials in the truncated theory better approximate the effective potential of the regular black hole (see Figure \ref{fig:D6potentials}). Nevertheless, the the convergence is again governed by the convergence of Hawking temperature, which is faster in this case (see Figure \ref{fig:THdev}).

We observe that the effective potentials for the truncated and full theories (see Figures \ref{fig:D5Haywardpotentials}~and~\ref{fig:D6potentials}) are nearly identical throughout the spacetime, except for a small region near the event horizon. This type of deviation has a negligible effect on the fundamental quasinormal mode but is expected to cause increasing discrepancies for higher overtones, as demonstrated in \cite{Konoplya:2022pbc}. Since grey-body factors are classical scattering characteristics primarily determined by the behavior of the potential near its peak, they converge as rapidly as the fundamental mode. As shown in Figures \ref{fig:GBdevd5} and \ref{fig:GBdevd6}, even at second order, the deviation from the full theory remains within one per cent. In contrast, the Hawking temperature, being a quantity determined by the near-horizon geometry, converges more slowly in truncated theories. This slower convergence directly affects the accuracy of the computed energy emission rates, as the temperature enters exponentially in the thermal spectrum. Similarly, other non-classical phenomena that depend sensitively on the near-horizon structure, such as particle production rates and entropy corrections, must be also impacted by this delayed convergence.

\vspace{-24pt}

\vspace{-16pt}

Thus, although the convergence is clear, the lowest orders do not provide sufficient approximation if one studies the near-horizon (quantum) phenomena. For example, in order to keep the relative difference with the full theory within one per cent, for the Hayward black hole one needs to take into account at least 8 orders in curvature.

\section{Conclusions}\label{sec:conclusions}

In the present work, we have not undertaken a comprehensive analysis of Hawking radiation for the full range of regular black hole solutions and their truncated approximations. The underlying framework admits a broad class of regular black holes characterized by a large number of parameters, which warrants a dedicated study. Our primary focus here was on the convergence properties of the infinite series that generate regular black hole geometries. This is particularly relevant for assessing the validity of studies that consider black holes with various curvature expansions truncated at quadratic, quartic or higher orders \cite{Bonanno:2025ohx,Konoplya:2019hml,Konoplya:2020jgt,Grain:2005my,Myers:1988ze,Camanho:2011rj,Rizzo:2006uz,Lu:2015cqa}. A previous investigation of classical observables, such as the fundamental quasinormal mode~\cite{Konoplya:2024hfg}, demonstrated a remarkably rapid convergence, with the relative difference already reduced to about one percent at second order. However, quantities that are strongly dependent on the near-horizon geometry---such as the Hawking temperature and evaporation rates---do not exhibit such exceptional accuracy. In these cases, several terms in the expansion are typically required to achieve a reliable~description.

An important question that lies beyond the scope of our present study is the stability of grey-body factors under small static deformations of the background spacetime. In~\cite{Konoplya:2025ixm}, the grey-body factors of the Schwarzschild black hole deformed by localized bumps and dips---near the event horizon and in the far zone---were calculated under the constraint that the Hawking temperature remains unchanged. It was shown there that the grey-body factors remain largely unaffected unless the deformations are sufficiently large, or the frequency is very low, in which case the grey-body factors are already extremely small and do not contribute to Hawking radiation. We believe this observation is sufficiently general and should hold even in theories with higher curvature corrections. After all, the convergence of the Hawking energy emission rate depends primarily on the temperature, and any deformation that modifies the Hawking temperature must originate from the underlying theory itself, rather than from ad hoc modifications of the effective potential.

Our goal in this work was to understand the general behavior of the convergence of higher curvature corrections, and for this reason we restricted our analysis to a test scalar field. A natural extension of this study would involve incorporating the emission of Standard Model fields in order to obtain more realistic estimates of Hawking radiation. Nevertheless, we expect that the qualitative features observed here will remain valid in such extended scenarios.

It is also worth mentioning that not all regular black hole metrics can be obtained via the above construction, as was shown in~\cite{Konoplya:2024kih}.
	

\section*{Author Contributions}
R.A.K.: conceptualization, methodology, software, visualization, writing; A.Z.: conceptualization, methodology, software, visualization, writing. All authors have read and agreed to the published version of the manuscript.

\section*{Funding}
This research received no external funding.

\section*{Institutional Review Board Statement}
Not applicable.

\section*{Informed Consent Statement}
Not applicable.

\section*{Data Availability Statement}
Not applicable.

\section*{Acknowledgments}
A. Z. would like to acknowledge Institute of Physics of Silesian University in Opava for support of his visit.

\section*{Conflicts of Interest}
The authors declare no conflict of interest.

\small
\bibliographystyle{scilight}


\end{document}